\documentclass[12pt]{jpconf}
\pdfoutput=1
\usepackage{times}
\usepackage{epsfig}
\usepackage{amsmath}
\usepackage{amsfonts}
\usepackage{color}

\usepackage{footmisc}
\renewcommand{\thefootnote}{\fnsymbol{footnote}}
\def\beq{\begin{equation}}
\def\eeq{\end{equation}}
\def\bea{\begin{eqnarray}}
\def\eea{\end{eqnarray}}
\def\beqn{\begin{eqnarray}} \def\eeqn{\end{eqnarray}}
\def\beeq{\begin{eqnarray}}
\def\eeeq{\end{eqnarray}}

\def\ep{\epsilon}

\def\Eq#1{Eq.~(\ref{#1})}

\def\bp{p\hspace{-.42em}/\hspace{-.07em}}
\def\bq{q\hspace{-.42em}/\hspace{-.07em}}

\def\td#1{\tilde{\delta}\left(#1\right)}

\def\qb{\mathbf{q}}

\begin{document}

\title{QED and QCD self-energy corrections\\ through the loop-tree duality}

\author{N. Selomit Ram\'irez-Uribe $^{1,2}$\footnote[1]{\hspace{1mm}Speaker} , Roger J. Hern\'andez-Pinto $^1$ and Germ\'an Rodrigo $^3$} 

\address{$^1$ Facultad de Ciencias F\'isico-Matem\'aticas, Universidad Aut\'onoma de Sinaloa, Avenida de las Am\'ericas y Boulevard Universitarios, Ciudad Universitaria, CP 80000, Culiac\'an, Sinaloa, M\'exico}

\address{$^2$ Facultad de Ciencias de la Tierra y el Espacio, Universidad Aut\'onoma de Sinaloa, Avenida de las Am\'ericas y Boulevard Universitarios, Ciudad Universitaria, CP 80000, Culiac\'an, Sinaloa, M\'exico}

\address{$^3$ Instituto de F\'{\i}sica Corpuscular, Universitat de Val\`{e}ncia-Consejo Superior de Investigaciones Cient\'{\i}ficas, Parc Cient\'{\i}fic, E-46980 Paterna, Valencia, Spain}

\ead{selomitru@uas.edu.mx, roger@uas.edu.mx, german.rodrigo@csic.es}

\begin{abstract}
The loop-tree duality (LTD) theorem establishes that loop contributions to scattering amplitudes can be computed through dual integrals, which are build from single cuts of the virtual diagrams. In order to build a complete LTD representation of a cross section and to achieve a local cancellation of singularities, it is crucial to include the renormalized self-energy corrections in an unintegrated form. In this document, we calculate the scalar functions related to the self-energy corrections in QED and QCD in the LTD formalism and extract explicitly their UV behaviour.
\end{abstract}

\section{Introduction}
\label{sec:intro}
The standard approach to perform perturbative calculations in QCD relies in the application of the subtraction formalism. There are diverse alternatives of the subtraction method at NLO and beyond ~\cite{Kunszt:1992tn, Frixione:1995ms, Catani:1996jh, Catani:1996vz, GehrmannDeRidder:2005cm, Catani:2007vq, Czakon:2010td, Bolzoni:2010bt, DelDuca:2015zqa, Boughezal:2015dva, Gaunt:2015pea, Compa:2014cp}, that involve the treatment of real and virtual contributions separately. From a computational point of view, handling real and virtual contributions separately might not be efficient enough for multi-leg and multi-loop processes. This is due to the fact that the final-state phase-space (PS) of the different contributions implies different number of external particles and loop momenta. For instance, at NLO, virtual corrections with Born kinematics have to be combined with real contributions with one additional final-state particle. The infrared (IR) counter-terms in the subtraction formalism have to be local in the real PS, and analytically integrable over the extra-radiation in order to properly cancel the divergent structure present in the virtual corrections. Building these counter-terms represents a challenge and introduces a potential bottleneck to efficiently carry out the IR subtraction for multi-leg multi-loop processes.

With the purpose of building renormalized quantities, we work the renormalization constants with the application of the LTD ~\cite{ Rodrigo:2008fp, Bierenbaum:2010cy, Bierenbaum:2012th, Bierenbaum:2013nja, Buchta:2015xda, Buchta:2015jea, Buchta:2015lva, Buchta:2015wna, Hernandez-Pinto:2015ysa, Sborlini:2016fcc}. The LTD theorem establishes a direct connection among loop and phase-space integrals. Then, loop scattering amplitudes can be expressed as a sum of PS integrals, called dual integrals. Dual integrals and real-radiation contributions exhibit a similar structure and in consequence, the divergent behaviour of both contributions is cancelled at integrand level, thus, rendering integrals finite. A remarkable implication of this formalism, is the possibility of carrying out purely four-dimensional implementations for any observable at NLO and higher orders.
\\
\\
\\
In this document we compute the scalar functions for the photon, electron, quark and gluon self-energies, within the LTD framework. Furthermore, we focus on the singular behaviour of the dual-integrals and the location of the singularities in the loop momentum space.

\section{Review of the loop-tree duality}
\label{sec:notationINTRO}
\begin{figure}[htb]
\begin{center}
\includegraphics[width=0.3\textwidth]{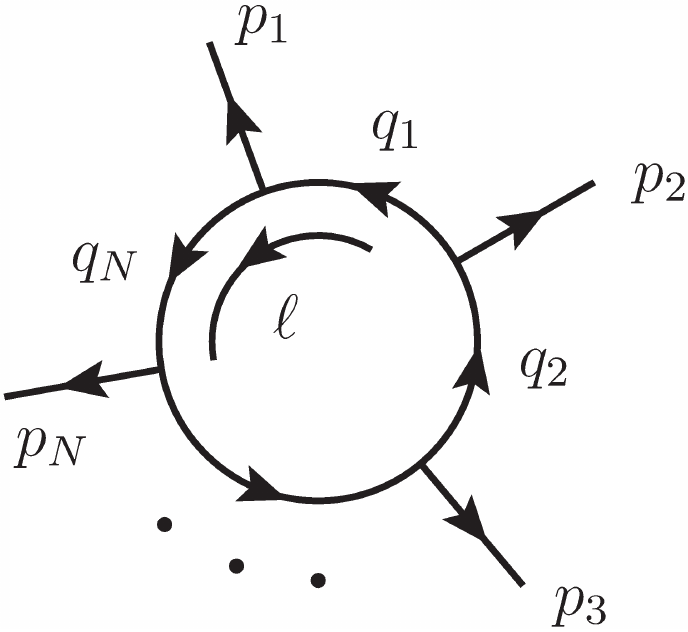}
\caption{Generic one-loop topology with $N$ external legs. 
\label{fig:OneLoopScalar}}
\end{center}
\end{figure}

In this section, we review the main ideas behind the LTD method. Using the LTD ~\cite{Catani:2008xa}, loop contributions of scattering amplitudes in any relativistic, local and unitary quantum field theory can be computed through dual integrals, which are build from single cuts of the virtual diagrams. Let's consider a generic $N$-particle scalar one-loop integral, i.e. 
\beq
L^{(1)}(p_1, \dots, p_N) = \int_{\ell} \, 
\prod_{i \in \alpha_1} \,G_F(q_i)~,
\label{oneloop}
\eeq 
over Feynman propagators $G_F(q_i) = (q_i^2-m_i^2+\imath 0)^{-1}$, whose most general topology is shown in Fig.~\ref{fig:OneLoopScalar}. The corresponding dual representation of \Eq{oneloop} consists in the sum of $N$ dual integrals:
\beq
L^{(1)}(p_1, \dots, p_N) 
= - \sum_{i\in \alpha_1} \, \int_{\ell} \, \td{q_i} \,
\prod_{j \in \alpha_1, \, j\neq i} \,G_D(q_i;q_j)~, 
\label{oneloopduality}
\eeq 
where 
\beq
G_D(q_i;q_j) = \frac{1}{q_j^2 -m_j^2 - \imath 0 \, \eta \cdot k_{ji}}\, ,
\eeq
are dual propagators, $k_{ji}=q_j-q_i$, and $i,j \in \alpha_1 = \{1,2,\ldots N\}$ label 
the available internal lines. In~\Eq{oneloop} and~\Eq{oneloopduality},
the masses and momenta of the internal lines are denoted $m_i$ and 
$q_{i,\mu} = (q_{i,0},\mathbf{q}_i)$, respectively, 
where $q_{i,0}$ is the energy and $\qb_{i}$ are 
the spatial components. In terms of the loop momentum $\ell$
and the outgoing momenta of the external particles $p_i$, the internal momenta are 
defined as 
\beq
q_{i} = \ell + k_i \ \ \ , \ \ \ k_{i} = \sum_{j=1}^i p_{j} \, ,
\eeq
together with the constraint $k_{N} = 0$ imposed by momentum conservation. 

On the other hand, the $d$-dimensional loop measure is given by
\beq
\int_{\ell} \bullet 
= - \imath \mu^{4-d} \int \frac{d^d \ell}{(2\pi)^{d}} \bullet~ \, ,
\eeq
and
\beq
\td{q_i} \equiv 2 \pi \, \imath \, \theta(q_{i,0}) \, \delta(q_i^2-m_i^2) \, 
\label{eq:tddefinition}
\eeq
used in \Eq{oneloopduality} set the internal lines on-shell. It is worth mentioning that, the presence of the Heaviside function restricts the integration domain to the positive energy region (i.e. $q_{i,0}>0$). Since LTD is derived through the application of the Cauchy's residue theorem, the remaining $d-1$ dimensional integration is performed over the forward on-shell hyperboloids defined by the solution of $G_F(q_i)^{-1}=0$ with $q_{i,0}>0$. Notice that these on-shell hyperboloids degenerate to light-cones when internal particles are massless.

The dual representation shown in \Eq{oneloopduality} is built by adding all possible single-cuts from the original loop diagram. In this procedure, the propagator associated with the on-shell line is replaced by \Eq{eq:tddefinition} where the remaining uncut Feynman propagators are promoted to dual ones. The introduction of dual propagators modifies the $\imath 0$-prescription since it depends on the sign of $\eta \cdot k_{ji}$, with $\eta$ a {\em future-like} vector, $\eta^2 \ge 0$, with positive definite energy $\eta_0 > 0$, and $k_{ji}$, which is independent of the loop momentum $\ell$ at one-loop. According to the derivation shown in Ref. \cite{Catani:2008xa}, $\eta$ is arbitrary so we can chose $\eta_\mu = (1,{\bf 0})$ to simplify the implementation. 

The difference between LTD and the Feynman Tree Theorem (FTT)~\cite{Feynman:1963ax,Feynman:1972mt}, where the loop integral is obtained after summing over all possible $m$-cuts, is codified in the dual prescription, where correlations coming from multiple cuts in the FTT are recovered in the LTD by considering only single-cuts with the modified $\imath 0$-prescription. In other words, having different prescriptions for each cut is a necessary condition for the consistency of the method. As discussed in Ref.~\cite{Buchta:2014dfa}, the integrand in~\Eq{oneloopduality} becomes singular at the intersection of forward on-shell hyperboloids (FF case),  and forward with backward ($q_{j,0}<0$) on-shell hyperboloids (FB intersections). On one hand, the FF singularities cancel each other among different dual contributions; the change of sign in the modified prescription is crucial to enable this behaviour. On the other hand, singularities associated with FB intersections remain constrained to a compact region of the loop three-momentum space and are easily reinterpreted in terms of causality. 

From a physical point of view, FB singularities take place when the on-shell virtual particle interacts with another on-shell virtual particle after the emission of outgoing on-shell radiation. The direction of the internal momentum flow establishes a natural causal ordering, and this interpretation is consistent with the Cutkosky rule. In fact, the total energy of the emitted particles, which is equal to $q_{i,0}-|q_{j,0}|$, has to be positive. Together with the positive energy constraint imposed by the delta distribution in~\Eq{eq:tddefinition}, it restricts the possible kinematical configurations compatible with a sequential decay of on-shell physical particles.

Besides the mathematical understanding of the origin of the singularities by using the LTD, one of the most promising applications is the possible computation of scattering processes at NLO directly in four space-time dimensions without Dimensional Regularization (DREG). It has been also shown in Ref.\cite{Sborlini:2015uia} that one of the ingredients in order to build the 4D integrals are the renormalization  constants. In fact, even if some integrals vanish at integral level, they are important for the local cancellation of singularities and the definition of renormalizad quantities in the LTD formalism. Therefore, in the next sections we compute the QED and QCD self-energy corrections by using the LTD theorem.
\section{Photon and electron self-energies in QED}
In the framework of QED, we need to compute the renormalization constants of the photon and the electron self-energies. In this handwritten, we will work in the massless approximation, labelling the internal lines as $q_1=\ell$, $q_2=\ell +p$ and considering $p^2\not= 0$.
\subsection{Photon self-energy}
\label{sec:photonprop}
The one-loop photon self-energy is given by
\beqn
\imath \, \Pi_{\mu\nu}(p^2) = -  \int_\ell {\mbox{Tr}} \left[ \imath e_0 \gamma_{\mu} \left( \imath \frac{\bq_1+m_0}{q_1^2-m_0^2+\imath 0}\right)\imath e_0\gamma_{\nu} \left( \imath \frac{\bq_2+m_0}{q_2^2-m_0^2+\imath 0}\right)\right]\, ,\label{F.photon}
\eeqn
where due to gauge invariance
\beqn
\Pi_{\mu\nu}(p^2)=(p^2\, g_{\mu\nu}-p_\mu p_\nu)\Pi(p^2)\, .\label{trans.photon}
\eeqn
From \Eq{F.photon} and \Eq{trans.photon} and taking the massless approximation,
\beqn
\imath(p^2\, g_{\mu\nu}-p_\mu p_\nu)\Pi(p^2)=- e_0^2 \int_\ell \frac{\mbox{Tr} \left[\gamma_{\mu}\, \bq_1\, \gamma_{\nu}\, \bq_2 \right]}{(q_1^2+\imath 0)\, (q_2^2+\imath 0)}\, .\label{photon}
\eeqn
Contracting \Eq{photon} with the metric tensor in d-dimensions we can extract the scalar function as,
\beqn
\Pi(p^2)
= \frac{8\, (1-\ep)\, e_0^2}{(3-2\ep)\, p^2}\int_{\ell}  \, \frac{q_1 \cdot q_2}{(q_1^2+\imath 0)\,  (q_2^2+\imath 0)}\, .\label{photon}
\eeqn
Applying the LTD to the integral in \Eq{photon}\, ,
\beqn
 \Pi(p^2) 
&=& -\frac{8\, (1-\ep)\, e_0^2}{(3-2\ep)\, p^2}\, \left[ \int_{\ell} \tilde\delta(q_1) \frac{p\cdot q_1}{2p\cdot q_1+p^2 -\imath 0}   - \int_{\ell}  \tilde\delta(q_2) \frac{p\cdot q_2}{-2p\cdot q_2+p^2+\imath 0}  \right]\, ,
\eeqn 
and taking the parametrization 
\beqn
p^{\mu} &=& (p_0,{\bf 0})\, , \nonumber\\
q_i^{\mu} &=& p_0 \xi_{i,0} \left( 1, 2 \sqrt{v_i(1-v_i)} {\bf e}_{i,\perp},1-2v_i \right)\, ,\nonumber \\ \label{parameters}
\eeqn
\beqn
\hspace{-2 mm}\Pi(p^2)
= -\frac{2^5\, (1-\ep)\, e_0^2}{(3-2\ep)}\,
\left[ \int d[\xi_{1,0}]d[v_1] \frac{\xi^2_{1,0}}{1+ 2\xi_{1,0}-\imath 0}  - \int d[\xi_{2,0}]d[v_2] \frac{\xi^2_{2,0}}{1-2\xi_{2,0}+\imath 0} \right]\, ,\label{photonLTD}
\eeqn
where
\beqn
d[\xi_{i,0}] = \frac{(4\pi)^{\ep-2}}{\Gamma(1-\ep)} \left(\frac{4p^2}{\mu^2} \right)^{-\ep} \xi_{i,0}^{-2\ep} d\xi_{i,0} \quad {\rm and } \quad d[v_i] = (v_i(1-v_i))^{-\ep} dv_i \, .\label{diferencial}
\eeqn
The parameters $\xi_{i,0}$ and $v_i$ describe the energy and polar angle of the loop momentum respectively, defined in the regions $\xi_{i,0} \in [ 0, \infty) $ and $v_i \in [0,1]$.
In order to explore the singular behaviour of $\Pi(p^2)$, we integrate \Eq{photonLTD} over the regions $i)\, \xi_i\in[0,\omega]$ and $ii)\, \xi_i\in[\omega,\infty)$.
After integration, we found:
	\beqn	
	\Pi(p^2;\xi_i \leq \omega)=-\frac{4\, e_0^2\, S_\ep}{3} \left[4\, \omega^2 +\log(1+2\, \omega -\imath 0)+\log(1-2\, \omega +\imath 0)\right]+O(\ep)\, \label{ser.photon.0w}
	\eeqn
	and  
	\beqn
	\Pi(p^2;\xi_i \geq \omega)=-\frac{4\, e_0^2\, S_\ep}{3} \left[\frac{1}{\ep} -4\, \omega^2 -\log(1+2\, \omega -\imath 0)-\log(1-2\, \omega +\imath 0) +\frac{5}{3} \right]+O(\ep)\, ,\label{ser.photon.winf}
	\eeqn	
where $S_\epsilon=\frac{(4\pi)^{-2+\epsilon}}{\Gamma(1-\epsilon)\mu^{-2\ep}}(-p^2-\imath 0)^{-\ep}$.
From these results, it is clear that IR divergences are absent. Besides, $\Pi(p^2;\xi_i \leq \omega)$ is finite while the ultraviolet (UV) divergence is located fully in $\Pi(p^2;\xi_i \geq \omega)$ . Since $\Pi(p^2;\xi_i \leq \omega)$ is IR finite, it admits a four dimensional integrand realization given by,
\beqn
\Pi(p^2;\xi_i \leq \omega)=-\frac{e_{0}^{2}}{6\pi^2}\int_0^1 dv_{1}\int_0^\omega \frac{d\xi_{1,0}\, 4\xi_{1,0}^2}{1+2\xi_{1,0}-\imath 0}\left[1+\log \left(\xi_{1,0}^2 \right) \left(v_1\delta \left(v_1\right) +\left(
1-v_1\right) \delta \left(1-v_1\right) \right) \right] \nonumber\\
+\frac{e_0^2}{6\pi^2} \int_0^1 dv_2 \int_0^\omega \frac{d\xi_{2,0}\, 4\xi_{2,0}^2}{1-2\xi_{2,0}+\imath 0}\left[1+\log \left(\xi_{2,0}^2\right) \left(v_2\delta \left(v_2\right) +\left(1-v_2\right) \delta \left( 1-v_2\right) \right) \right] ,\label{finite.photon}
\eeqn
where the presence of the logarithmic terms are originated from the fact that we are using different coordinate systems for each dual integral.

The previous results shows one of the central implications of the use of the LTD theorem, resides in the location of singularities in physical regions. Hence, segmenting the integration domain, it is possible to find regions where the $\ep$-parameter is no longer needed. It is also worth to highlight that in \Eq{finite.photon} is showing explicitly the cut dependency, this fact will be relevant when a physical process with photons is calculated at NLO accuracy.
\subsection{Electron self-energy}
\label{sec:electronprop}
The one-loop electron self-energy is given by
\beqn
-\imath \Sigma(p) = \int_\ell  \imath e_0 \gamma^{\mu} \left( \imath \frac{\bq_2 +m}{q_2^2 - m_0^2+\imath 0} \right)\imath e_0 \gamma^{\nu} \frac{-\imath}{q_1^2+\imath 0} g_{\mu\nu}\, ,\label{F.electron}
\eeqn
where
\beqn
\Sigma(p) = \bp \Sigma_V(p^2) -m \Sigma_S(p^2)\, .\label{trans.electron}
\eeqn
Considering \Eq{F.electron}, \Eq{trans.electron} and the massless case, we get,
\beqn
 -\imath\, \bp\, \Sigma_V(p^2) = -e_0^2 \int_\ell \frac{\gamma^{\mu} \bq_2\gamma_{\mu} }{(q_1^2+\imath 0)(q_2^2+\imath 0)}\, . \label{electron.case}   
\eeqn
Extracting the scalar function $\Sigma_V(p^2)$, we find,
\beqn
\Sigma_V(p^2)  
&=& - 2\, (1-\ep)\, e_0^2 \int_{\ell}\, \frac{1}{(q_1^2+\imath 0)\, (q_2^2+\imath 0)} \left(1  +\frac{p\cdot q_2}{p^2}\right)\, . \label{electron.LTD}
\eeqn
Then, applying the LTD,
\beqn
\Sigma_V(p^2) 
&=&   2 (1-\ep) \, e_0^2 \left( \int_{\ell}     \frac{\tilde{\delta}(q_1) }{2p\cdot q_1 +p^2-\imath 0} \left(1  +\frac{p\cdot q_1}{p^2}\right) + \frac{1}{p^2}\int_{\ell}     \frac{\tilde{\delta}(q_2) \, p\cdot q_2 }{-2p\cdot q_2 +p^2+\imath 0}  \right)\, ,
\eeqn
working with the parametrization defined in \Eq{parameters}, $\Sigma_V(p^2)$ becomes,
\beqn
\hspace{-5 mm}\Sigma_V(p^2)  
&=&   2 (1-\ep) \, e_0^2 \left( \int d[\xi_{1,0}]d[v_1]     \frac{4\xi_{1,0}\left(1  + \xi_{1,0}\right) }{1+2\xi_{1,0}-\imath 0}  +\int d[\xi_{2,0}]d[v_2]     \frac{4\xi_{2,0}^2 }{1-2\xi_{2,0}+\imath 0}  \right) \, .
\eeqn
To explore the behaviour of singularities from $\Sigma_V(p^2)$, we segment the integration region similar to the photon propagator case. Computing the integrals in both regions, we find, 
	\beqn	
	\Sigma_V(p^2;\xi_i \leq \omega)=e_{0}^{2}\, S_\ep \left[-\log (1+2\omega -\imath 0) -\log ( 1-2\omega +\imath 0) \right]+O\left( \epsilon \right) \, ,\label{ser.electron.0w}
	\eeqn	
	\beqn
	\Sigma_V(p^2;\xi_i \geq \omega)=e_{0}^{2}\, S_\ep \left[-\frac{1}{\ep} +\log (1+2\omega -\imath 0) +\log ( 1-2\omega +\imath 0)-1 \right]+O(\epsilon ) \, . \label{ser.electron.winf}
	\eeqn

Based on these we find that $\Sigma_V(p^2)$ has not IR divergence and similar conclusions to the photon propagator case can be inferred. Since $\Sigma_V(p^2;\xi_i \leq \omega)$ is finite, we found that,
\beqn
\Sigma_V(p^2;\xi_i \leq \omega)=\frac{e_{0}^{2}}{8\pi ^{2}}\int_{0}^{1}dv_{1}\int_{0}^{\omega}\, d\xi_{1,0}\frac{4\xi _{1,0}\left( 1+\xi_{1,0}\right) }{1+2\xi_{1,0}-\imath 0}\left[
1+\log \left( \xi_{1,0}^{2}\right) \left[v_{1}\delta \left(v_{1}\right)
+\left(1-v_{1}\right) \delta \left(1-v_{1}\right) \right] \right] \nonumber\\
+\frac{e_{0}^{2}}{8\pi ^{2}}\int_{0}^{1}dv_{2}\int_{0}^{\omega}\, d\xi
_{2,0} \frac{4\xi _{2,0}^{2}}{1-2\xi _{2,0}+\imath 0}\left[ 1+\log \left( \xi_{2,0}^{2}\right) \left[ v_{2}\delta \left( v_{2}\right) +\left(
1-v_{2}\right) \delta \left(1-v_{2}\right) \right] \right]\, ,
\eeqn
is a four dimensional representation of it. It is worth mentioning that the massive case has been studied in Ref.\cite{Sborlini:2016fcj}, but in this work we are only focused in the one-loop self-energies corrections to massless propagators.
\section{Quark and gluon self-energies in QCD}
In this section we explicitly compute the quark and gluon self-energy corrections at one-loop in QCD.
\subsection{Quark self-energy}
\label{sec:quarkprop}

One loop correction to the quark self-energy has a similar structure to the electron propagator in QED. The one-loop quark self-energy differs from the one-loop electron self-energy, only by the replacement of $e_0$ by $g_0$ and the inclusion of the colour factor $C_F$. 

Then, we have that the scalar function $\Sigma_V(p^2)$ is finite in the region $[0, \omega)$ and has a pole related to the presence of UV divergence at $[\omega, \infty]$. As in the electron case, there is a four dimensional realization given by
\beqn
\Sigma_V(p^2;\xi_i \leq \omega)&=&\frac{C_F\, g_{0}^{2}}{8\pi ^{2}}\int_{0}^{1}dv_{1}\int_{0}^{\omega}\, d\xi_{1,0}\frac{4\xi _{1,0}\left( 1+\xi_{1,0}\right) }{1+2\xi_{1,0}-\imath 0}\left[
1+\log \left( \xi_{1,0}^{2}\right) \left[v_{1}\delta \left(v_{1}\right)+\left(1-v_{1}\right) \delta \left(1-v_{1}\right) \right] \right] \nonumber\\
&+&\frac{C_F\, g_{0}^{2}}{8\pi ^{2}}\int_{0}^{1}dv_{2}\int_{0}^{\omega}\, d\xi
_{2,0} \frac{4\xi _{2,0}^{2}}{1-2\xi _{2,0}+\imath 0}\left[ 1+\log \left( \xi_{2,0}^{2}\right) \left[ v_{2}\delta \left( v_{2}\right) +\left(1-v_{2}\right) \delta \left(1-v_{2}\right) \right] \right]. \nonumber \\
\eeqn

\subsection{Gluon self-energy}
\label{sec:gluonprop}
We analyse now the one-loop QCD correction to the gluon self-energy (in the Feynman gauge), which receives three different contributions, the quark, the gluon and the ghost-loop. For the gluon self-energy, the one-loop scalar function can be written as:
\beqn
\Pi(p^2)=\Pi_q(p^2)+\Pi_1(p^2)+\Pi_2(p^2)\, ,
\eeqn
where $\Pi_q(p^2)$, $\Pi_1(p^2)$ and $\Pi_2(p^2)$ are the corresponding scalar functions belonging to the quark, the gluon and the ghost-loop, respectively.

The quark-loop contribution differs from the one-loop photon self-energy only by the replacement of $e_0$ by $g_0$ and the inclusion of the factor $n_f$, representing the number of fermions in the loop, therefore;
\beqn
\Pi_q(p^2)  
&=& \frac{8\, n_f\, (1-\ep)\, g_0^2}{(3-2\ep)\, p^2}\int_{\ell}  \, \frac{q_1 \cdot q_2}{(q_1^2+\imath 0)\,  (q_2^2+\imath 0)}\, .
\eeqn
Now, the contribution coming from the gluon-loop is
\beqn
\Pi_1(p^2) 
&=& - \frac{\imath\, C_A g_0^2}{2(3-2\ep)\, p^2} \int_\ell \frac{N}{(q_1^2+\imath 0)\, (q_2^2+\imath 0)}\, , \label{gluon.LTD}
\eeqn
where the numerator is
\beqn
N&=&d((q_1+q_2)^2+ (2q_1-q_2)^2+(2q_2-q_1)^2)\nonumber\\
&-&2(q_1+q_2)\cdot(2q_1-q_2)-2q_2\cdot(2q_2-q_1) +2(2q_2-q_1)\cdot(2q_1-q_2)\, .
\eeqn
Applying the LTD to the integral in \Eq{gluon.LTD} , we find,
\beqn
\Pi_1(p^2) 
=-3\, C_A \, g_0^2 \left(  \int_{\ell} \frac{\tilde{\delta}(q_1) }{p^2+2p\cdot q_1-\imath 0}\left(1+\frac{p\cdot q_1}{p^2} \right) + \int_{\ell}\frac{\tilde{\delta}(q_2)}{p^2-2p\cdot q_2 +\imath 0}\left(1-\frac{p\cdot q_1}{p^2} \right) \right) \, .
\eeqn
Using the same parametrization, we find,
\beqn
\Pi_1(p^2)
&=& -\frac{12\, C_A \, g_0^2}{\Gamma(2-2\ep)} \left(  \int d[v_1]\, d[\xi_{1,0}] \frac{\xi_{1,0}\, (1+\xi_{1,0})}{ 1+2\xi_{1,0}-\imath 0} + \int d[v_2]\, d[\xi_{2,0}]\frac{\xi_{2,0}\, (1-\xi_{2,0})}{1-\xi_{2,0} +\imath 0} \right)\, .\label{scalar.gluon}
\eeqn

Similar to the QED results, we find after integration
	\beqn	
	\Pi_1(p^2;\xi_i \leq \omega)=\frac{C_{A}\, g_{0}^{2}\, S_\ep}{3}  \left[ 4\omega ^{2}-\log \left(1+2\omega -\imath 0\right) -\log \left( 1-2\omega +\imath 0\right)\right]+O(\epsilon )\, ,
	\eeqn
	\beqn
	\Pi_1(p^2;\xi_i \geq \omega)=
	\frac{C_{A}\, g_{0}^{2}\, S_\ep}{3} \left[ -\frac{1}{\epsilon }-4\omega ^{2}+\log \left(1+2\omega -\imath 0\right) +\log \left( 1-2\omega +\imath 0\right)-\frac{4}{3} \right] +O\left( \epsilon \right). 
	\eeqn
Therefore, the four dimensional representation of $\Pi_1(p^2;\xi_i \leq \omega)$ is given by,
\beqn
\hspace{-1.2 mm} \Pi_1(p^2;\xi_i \leq \omega)&=&
-\frac{9 C_A g_{0}^{2}}{4\pi^{2}}\int_{0}^{1}dv_{1} \int_{0}^{\omega }\frac{d\xi _{1,0}\xi _{1,0}\left( 1+\xi _{1,0}\right) }{1+2\xi _{1,0}-\imath 0}[1+\log \left( \xi _{1,0}^{2}\right) \left[ v_{1}\delta \left( v_{1}\right) +\left( 1-v_{1}\right) \delta \left( 1-v_{1}\right) \right] ] \nonumber \\
&-&\frac{9 C_A g_{0}^{2}}{4\pi^{2}}\int_{0}^{1}dv_{2}\int_{0}^{\omega }\frac{d\xi _{2,0}\xi _{2,0}\left(1-\xi _{2,0}\right) }{1-2\xi _{2,0}+\imath 0}[1+\log \left( \xi _{2,0}^{2}\right) \left[ v_{2}\delta \left( v_{2}\right)+\left( 1-v_{2}\right) \delta \left( 1-v_{2}\right) \right]]\, .\label{4d.gluon}\nonumber \\
\eeqn
The last remaining piece is due to the ghost-loop, the interested integral is given by,
\beqn	
\Pi_2(p^2)
&=& \frac{\imath\, C_A\, g_0^2}{(3-2\ep)p^2} \int_\ell \frac{q_1 \cdot q_2 }{(q_1^2+\imath 0)\, (q_2^2+\imath 0)} 
\eeqn
applying LTD to the previous expression,
\beqn
\Pi_2(p^2)
&=& \frac{C_A  \,g_0^2}{(3-2\ep)p^2} \left( \int_{\ell}\frac{\tilde{\delta}(q_1)q_1 \cdot p}{p^2+2p\cdot q_1-\imath 0} -\int_{\ell}\frac{\tilde{\delta}(q_2)p \cdot q_2}{p^2-2p\cdot q_2 +\imath 0} \right)    
\, ,
\eeqn
Following the precedent ideas, we find that the integral is decomposed as
	\beqn	
	\Pi_2(p^2;\xi_i \leq \omega)=\frac{C_{A}g_{0}^{2}\, S_\ep}{3}  \left[4\omega ^{2}+\log \left( 1+2\omega -\imath 0\right)+\log \left( 1-2\omega +\imath 0\right) \right]+O\left(\epsilon \right)\, ,
	\eeqn
	\beqn
	\Pi_2(p^2;\xi_i \geq \omega)=\frac{C_{A}g_{0}^{2}\, S_\ep}{3} \left[ \frac{1}{\epsilon }-4\omega ^{2}-\log \left( 1+2\omega -\imath 0\right)-\log \left( 1-2\omega +\imath 0\right)+2 \right]+O\left( \epsilon \right) \, .
	\eeqn
Finally, we find that the four dimensional representation of $\Pi_2(p^2;\xi_i \leq \omega)$ is given by
\beqn
\Pi_2(p^2;\xi_i \leq \omega)&=&\frac{C_{A}g_{0}^{2}}{48\pi ^{2}}\int_{0}^{1}dv_{1}\int_{0}^{\omega }
\frac{d\xi _{1,0}4\xi _{1,0}^{2}}{1+2\xi _{1,0}-\imath 0}\left[ 1+\log
\left( \xi _{1,0}^{2}\right) \left[ v_{1}\delta \left( v_{1}\right) +\left(1-v_{1}\right) \delta \left( 1-v_{1}\right) \right] \right] \nonumber\\
&+&\frac{C_{A}g_{0}^{2}}{48\pi ^{2}}\int_{0}^{1}dv_{2}\int_{0}^{\omega } \frac{d\xi _{2,0}4\xi _{2,0}^{2}}{1-2\xi _{2,0}+\imath 0}\left[ 1+\log\left( \xi _{2,0}^{2}\right) \left[ v_{2}\delta \left( v_{2}\right) +\left(1-v_{2}\right) \delta \left( 1-v_{2}\right) \right] \right] \label{4d.ghost}
\eeqn
Gathering the four dimensional representations of the three different contributions to the gluon self-energy, the quark, the gluon and the ghost loop, we have that the four dimensional realization of $\Pi(p^2;\xi_i \leq \omega$) is given by,
\beqn
\Pi(p^2;\xi_i \leq \omega)=\Pi_q(p^2;\xi_i \leq \omega)+\Pi_1(p^2;\xi_i \leq \omega)+\Pi_2(p^2;\xi_i \leq \omega)\, .
\eeqn
\section{Conclusions}
\label{sec:conclusions}
In this paper we have shown that working with the LTD method the scalars functions from the one-loop self energies present only simple poles related to their UV-singular behaviour. Therefore a purely four dimensional representation could be always found for a region belonging to the $\omega$-cut in the energy component. The possibility to obtain a four dimensional realization implies a major improvement in the computation of higher-order corrections in QFT, besides, allows a better understanding of the mathematical structures behind scattering amplitudes.

\section{Acknowledgment}
This work is supported by CONACyT, Mexico, PROFAPI 2015 grant number 121 and DPyC, by the Spanish Government and ERDF funds from European Commission (Grants No. FPA2014-53631-C2-1-P and SEV-2014-0398) and by Consejo Superior de Investigaciones Cient\'ificas (Grant No. PIE-201750E021)



\section{References}

\begin{thebibliography}{90}



\bibitem{Kunszt:1992tn}
  Z.~Kunszt and D.~E.~Soper,
  ``Calculation of jet cross-sections in hadron collisions at order alpha-s**3,''
  Phys.\ Rev.\ D {\bf 46} (1992) 192.
\bibitem{Frixione:1995ms}
  S.~Frixione, Z.~Kunszt and A.~Signer,
  ``Three jet cross-sections to next-to-leading order,''
  Nucl.\ Phys.\ B {\bf 467} (1996) 399
  [hep-ph/9512328].
\bibitem{Catani:1996jh}
  S.~Catani and M.~H.~Seymour,
  ``The Dipole formalism for the calculation of QCD jet cross-sections at next-to-leading order,''
  Phys.\ Lett.\ B {\bf 378} (1996) 287
  [hep-ph/9602277].
\bibitem{Catani:1996vz}
  S.~Catani and M.~H.~Seymour,
  ``A General algorithm for calculating jet cross-sections in NLO QCD,''
  Nucl.\ Phys.\ B {\bf 485} (1997) 291
   [Nucl.\ Phys.\ B {\bf 510} (1998) 503]
  [hep-ph/9605323].
\bibitem{GehrmannDeRidder:2005cm}
  A.~Gehrmann-De Ridder, T.~Gehrmann and E.~W.~N.~Glover,
  ``Antenna subtraction at NNLO,''
  JHEP {\bf 0509} (2005) 056
  [hep-ph/0505111].
\bibitem{Catani:2007vq}
  S.~Catani and M.~Grazzini,
  ``An NNLO subtraction formalism in hadron collisions and its application to Higgs boson production at the LHC,''
  Phys.\ Rev.\ Lett.\  {\bf 98} (2007) 222002
  [hep-ph/0703012].

\bibitem{Czakon:2010td}
  M.~Czakon,
  ``A novel subtraction scheme for double-real radiation at NNLO,''
  Phys.\ Lett.\ B {\bf 693} (2010) 259
  [arXiv:1005.0274 [hep-ph]].
\bibitem{Bolzoni:2010bt}
  P.~Bolzoni, G.~Somogyi and Z.~Trocsanyi,
  ``A subtraction scheme for computing QCD jet cross sections at NNLO: integrating the iterated singly-unresolved subtraction terms,''
  JHEP {\bf 1101} (2011) 059
  [arXiv:1011.1909 [hep-ph]].

\bibitem{DelDuca:2015zqa}
  V.~Del Duca, C.~Duhr, G.~Somogyi, F.~Tramontano and Z.~Tr\'ocs\'anyi,
  ``Higgs boson decay into b-quarks at NNLO accuracy,''
  JHEP {\bf 1504} (2015) 036
  [arXiv:1501.07226 [hep-ph]].

\bibitem{Boughezal:2015dva}
  R.~Boughezal, C.~Focke, X.~Liu and F.~Petriello,
  ``$W$-boson production in association with a jet at next-to-next-to-leading order in perturbative QCD,''
  Phys.\ Rev.\ Lett.\  {\bf 115} (2015) 062002
  [arXiv:1504.02131 [hep-ph]].
  
\bibitem{Gaunt:2015pea}          
  J.~Gaunt, M.~Stahlhofen, F.~J.~Tackmann and J.~R.~Walsh,
  ``N-jettiness Subtractions for NNLO QCD Calculations,''
  JHEP {\bf 1509} (2015) 058
  [arXiv:1505.04794 [hep-ph]].

  

\bibitem{Compa:2014cp}
  R.~A.~Fazio, P.~Mastrolia, E.~Mirabella, W.~J.~Torres-Bobadilla,
  ``On the four-dimensional formulation of dimensionally regulated amplitudes,''
  Eur. Phys. J.C. {\bf 74} (2014) 3197


\bibitem{Rodrigo:2008fp}
  G.~Rodrigo, S.~Catani, T.~Gleisberg, F.~Krauss and J.~C.~Winter,
  ``From multileg loops to trees (by-passing Feynman's Tree Theorem),''
  Nucl.\ Phys.\ Proc.\ Suppl.\  {\bf 183} (2008) 262
  [arXiv:0807.0531 [hep-th]].

\bibitem{Bierenbaum:2010cy}
  I.~Bierenbaum, S.~Catani, P.~Draggiotis and G.~Rodrigo,
  ``A Tree-Loop Duality Relation at Two Loops and Beyond,''
  JHEP {\bf 1010} (2010) 073
  [arXiv:1007.0194 [hep-ph]].

\bibitem{Bierenbaum:2012th}
  I.~Bierenbaum, S.~Buchta, P.~Draggiotis, I.~Malamos and G.~Rodrigo,
  ``Tree-Loop Duality Relation beyond simple poles,''
  JHEP {\bf 1303} (2013) 025
  [arXiv:1211.5048 [hep-ph]].

\bibitem{Bierenbaum:2013nja}
  I.~Bierenbaum, P.~Draggiotis, S.~Buchta, G.~Chachamis, I.~Malamos and G.~Rodrigo,
  ``News on the loop--tree Duality,''
  Acta Phys.\ Polon.\ B {\bf 44} (2013) 2207.

\bibitem{Buchta:2015xda}
  S.~Buchta,
  ``Theoretical foundations and applications of the Loop-Tree Duality in Quantum Field Theories,''
  PhD thesis, Universitat de Val\`encia, 2015,
  arXiv:1509.07167 [hep-ph].

\bibitem{Buchta:2015jea}
  S.~Buchta, G.~Chachamis, P.~Draggiotis, I.~Malamos and G.~Rodrigo,
  ``Towards a Numerical Implementation of the Loop-Tree Duality Method,''
  Nucl.\ Part.\ Phys.\ Proc.\  {\bf 258-259} (2015) 33
  [arXiv:1509.07386 [hep-ph]].

\bibitem{Buchta:2015lva}
  S.~Buchta,
  ``First Numerical Implementation of the Loop-Tree Duality Method,''
  PoS EPS-HEP {\bf 2015} (2015) 430
  [arXiv:1510.04105 [hep-ph].

\bibitem{Buchta:2015wna}
  S.~Buchta, G.~Chachamis, P.~Draggiotis and G.~Rodrigo,
  ``Numerical implementation of the Loop-Tree Duality method,''
  arXiv:1510.00187 [hep-ph].

\bibitem{Hernandez-Pinto:2015ysa}
  R.~J.~Hern\'andez-Pinto, G.~F.~R.~Sborlini and G.~Rodrigo,
  ``Towards gauge theories in four dimensions,''
  JHEP {\bf 1602} (2016) 044
  [arXiv:1506.04617 [hep-ph]].

\bibitem{Sborlini:2016fcc}
  G.~F.~R.~Sborlini,
  ``Loop-tree duality and quantum field theory in four dimensions,''
  PoS RADCOR {\bf 2015} (2015) 082 
  [arXiv:1601.04634 [hep-ph]].
  \bibitem{Catani:2008xa}
  S.~Catani, T.~Gleisberg, F.~Krauss, G.~Rodrigo and J.~C.~Winter,
  ``From loops to trees by-passing Feynman's theorem,''
  JHEP {\bf 0809} (2008) 065
  [arXiv:0804.3170 [hep-ph]].



\bibitem{Feynman:1963ax}
  R.~P.~Feynman,
  ``Quantum theory of gravitation,''
  Acta Phys.\ Polon.\  {\bf 24} (1963) 697.
\bibitem{Feynman:1972mt}
  R.~P.~Feynman,
  ``Closed Loop And Tree Diagrams. (talk),''
  In *Brown, L.M. (ed.): Selected papers of Richard Feynman* 867-887

\bibitem{Buchta:2014dfa}
  S.~Buchta, G.~Chachamis, P.~Draggiotis, I.~Malamos and G.~Rodrigo,
  ``On the singular behaviour of scattering amplitudes in quantum field theory,''
  JHEP {\bf 1411} (2014) 014
  [arXiv:1405.7850 [hep-ph]].

\bibitem{Sborlini:2015uia}
  G.~F.~R.~Sborlini, R.~Hern\'andez-Pinto and G.~Rodrigo,
  ``From dimensional regularization to NLO computations in four dimensions,''
  PoS EPS-HEP {\bf 2015} (2015) 479
  [arXiv:1510.01079 [hep-ph]].

\bibitem{Sborlini:2016fcj}
  G.~F.~R.~Sborlini, Felix Driencourt-Mangin, German Rodrigo
  ``Four-dimensional unsubtraction with massive particles,''
  JHEP {\bf 1610} (2016) 162
  [arXiv:1608.01584 [hep-ph]].  
  

  
\end{thebibliography}
\end{document}